\documentclass[smallextended,numbook]{svjour3}
\smartqed

\usepackage[colorlinks=true, linkcolor=black!50!blue, urlcolor=blue, citecolor=blue, anchorcolor=blue]{hyperref}
\usepackage[font=small,labelfont=bf,margin=0mm,labelsep=period,tableposition=top]{caption}
\usepackage[a4paper,top=3cm,bottom=2.5cm,left=2.5cm,right=2.5cm,bindingoffset=0mm]{geometry}

\usepackage{framed}
\usepackage{graphicx,placeins}
\usepackage{float}
\usepackage{afterpage}
\usepackage{epsfig,cite}
\usepackage{amssymb}
\usepackage{amsmath}
\usepackage{dsfont}
\usepackage{multirow}
\usepackage{url}
\usepackage{xcolor}
\usepackage{float}
\usepackage{afterpage}

\usepackage{url}
\usepackage{hyperref}

\usepackage{booktabs}

\usepackage{tikz}
\usetikzlibrary{arrows}
\usepackage{tikz-3dplot}
\usepackage{enumitem}

\usepackage{color}
\usepackage{fancyvrb}

\DefineVerbatimEnvironment{Highlighting}{Verbatim}{commandchars=\\\{\}}
\newenvironment{Shaded}{}{}

\newcommand{\AttributeTok}[1]{\textcolor[rgb]{0.49,0.56,0.16}{#1}}

\newcommand{\CharTok}[1]{\textcolor[rgb]{0.25,0.44,0.63}{#1}}

\newcommand{\DecValTok}[1]{\textcolor[rgb]{0.25,0.63,0.44}{#1}}

\newcommand{\FunctionTok}[1]{\textcolor[rgb]{0.02,0.16,0.49}{#1}}

\newcommand{\KeywordTok}[1]{\textcolor[rgb]{0.00,0.44,0.13}{\textbf{#1}}}
\newcommand{\NormalTok}[1]{#1}

\newcommand{\StringTok}[1]{\textcolor[rgb]{0.25,0.44,0.63}{#1}}

\newcommand{\be}{\begin{equation}}
\newcommand{\ee}{\end{equation}}
\newcommand{\bea}{\begin{eqnarray}}
\newcommand{\eea}{\end{eqnarray}}
\newcommand{\bi}{\begin{itemize}}
\newcommand{\ei}{\end{itemize}}
\newcommand{\ben}{\begin{enumerate}}
\newcommand{\een}{\end{enumerate}}

\def\frac#1#2{{{#1}\over {#2}}}
\def\gsim{\mathrel{\rlap{\lower4pt\hbox{\hskip1pt$\sim$}}
    \raise1pt\hbox{$>$}}}         
\def\lsim{\mathrel{\rlap{\lower4pt\hbox{\hskip1pt$\sim$}}
    \raise1pt\hbox{$<$}}}         

\newcommand{\draft}[1]{}

\def\beq{\begin{equation}}
\def\eeq{\end{equation}}

\def\lapprox{\lower .7ex\hbox{$\;\stackrel{\textstyle <}{\sim}\;$}}
\def\gapprox{\lower .7ex\hbox{$\;\stackrel{\textstyle >}{\sim}\;$}}

\numberwithin{equation}{section}
\numberwithin{figure}{section}
\numberwithin{table}{section}

\usepackage{tabularx}
\newcolumntype{C}[1]{>{\centering\arraybackslash}p{#1}}

\usepackage{amsmath}
\usepackage{amsfonts}
\usepackage{amssymb}
\usepackage{dsfont}
\usepackage{pifont}
\usepackage{booktabs}
\usepackage{graphicx}
\usepackage{epstopdf}
\usepackage{epsfig}
\usepackage[framed]{ntheorem}
\usepackage{framed}
\usepackage{makeidx}
\usepackage{simplewick}
\usepackage{hyperref}
\usepackage{placeins}
\usepackage[font=small,labelfont=bf]{caption}

\begin{document}

\title{An open-source machine learning framework for global analyses of parton distributions}

\author{The NNPDF Collaboration: Richard D. Ball \and 
Stefano Carrazza\and Juan Cruz-Martinez\and Luigi Del Debbio\and Stefano Forte\and
Tommaso Giani\and Shayan Iranipour\and Zahari Kassabov\and Jose I. Latorre\and
Emanuele R. Nocera\and Rosalyn L. Pearson\and Juan Rojo\and Roy Stegeman\and
Christopher Schwan\and Maria Ubiali\and Cameron Voisey \and
Michael Wilson}
\institute{
Richard D. Ball \and Luigi Del Debbio \and Emanuele R. Nocera \and Rosalyn L. Pearson \and
Michael Wilson \at The Higgs Centre for
Theoretical Physics, University of Edinburgh,  JCMB, KB, Mayfield Rd,
Edinburgh EH9 3JZ, Scotland
\and
Stefano Carrazza\and Juan Cruz-Martinez \and Stefano Forte \and
Christopher Schwan \and Roy Stegeman \at Tif Lab, Dipartimento di Fisica, Universit\`a di Milano and
INFN, Sezione di Milano, Via Celoria 16, I-20133 Milano, Italy
\and
Shayan Iranipour \and Zahari Kassabov \and Maria Ubiali \at DAMTP,
University of Cambridge, Wilberforce Road, Cambridge, CB3 0WA, United
Kingdom
\and
 Jose I. Latorre \at Quantum Research Centre, Technology Innovation Institute, Abu Dhabi,
United Arab Emirates
\at Center for Quantum Technologies, National University of Singapore, Singapore 
\at  Qilimanjaro Quantum Tech, Barcelona, Spain
\and
Tommaso Giani \and Emanuele R. Nocera \and  Juan Rojo \at Department of Physics and Astronomy, VU Amsterdam,
NL-1081 HV Amsterdam
\at Nikhef Theory Group, Science Park 105, 1098 XG Amsterdam, The Netherlands
\and
Cameron Voisey \at Cavendish Laboratory, University of Cambridge, Cambridge, CB3 0HE, United Kingdom
}

\maketitle

\begin{abstract}
  We present the software framework underlying the NNPDF4.0 global
  determination of parton distribution functions (PDFs). The code is released
  under an open source licence and is accompanied by extensive documentation
  and examples.
  The code base is composed by a PDF fitting package, tools to
  handle experimental data and to efficiently compare it to theoretical
  predictions, and a versatile analysis framework.
  In addition to ensuring
  the reproducibility of the NNPDF4.0 (and subsequent) determination, the public
  release of the NNPDF fitting framework enables a number of phenomenological applications
  and the production of PDF fits under user-defined data and theory assumptions.\\[0.3cm]

  \noindent
  Edinburgh 2021/13\\
  Nikhef-2021-020\\
  TIF-UNIMI-2021-12
  
\end{abstract}

\tableofcontents

\clearpage
\section{Introduction}
\label{sec:introduction}

The success of the ambitious programme of the upcoming Run III at the LHC and its subsequent
High-Luminosity upgrade~\cite{Cepeda:2019klc,Azzi:2019yne} relies on achieving the highest possible
accuracy not only in the experimental measurements
but also in the corresponding
theoretical predictions.
A key component of the latter are the
parton distribution functions (PDFs),
which parametrize the quark and gluon
substructure of the colliding protons~\cite{Gao:2017yyd,Ethier:2020way}.
PDFs are dictated by non-perturbative QCD dynamics
and hence must be phenomenologically extracted  by matching a wide range
of experimental data with the corresponding
theoretical predictions.
The determination of PDFs and their uncertainties requires a robust statistical
framework which minimises unnecessary assumptions while implementing known
theoretical constraints such as QCD evolution, sum rules, positivity, and
integrability.

Recently, a new family of global PDF analyses has been presented by
the NNPDF Collaboration: NNPDF4.0~\cite{nnpdf40}.
This updated PDF determination framework
supersedes its predecessor NNPDF3.1~\cite{Ball:2017nwa}
by improving on all relevant aspects, from the experimental input
and theoretical constraints to the optimisation methodology and the validation of results.
As with previous NNPDF releases, the NNPDF4.0 PDFs are made publicly
available via the standard {\sc\small LHAPDF} interface~\cite{Buckley:2014ana}.
However, until now only the  outcome of the NNPDF fits (the {\sc\small LHAPDF}
interpolation grid files) was released,
while the code itself remained private.
This situation implied that the only option to produce tailored variants
of the NNPDF analyses was by requesting them to the developers, and further
that results were not reproducible by external parties.
Another limitation of private PDF codes is that
benchmarking studies, such as those carried out by the
PDF4LHC working group~\cite{Butterworth:2015oua,Rojo:2015acz},
become more convoluted due to the challenge in disentangling
the various components that determine the final outcome.
 
Motivated by this state of affairs, as well as by the principles of Open and FAIR~\cite{FAIR} (findable, accessible, interoperable and reusable) Science,
in this work we describe the public release of
the complete software framework~\cite{nnpdfcode}
underlying the NNPDF4.0 global determination
together with user-friendly examples and an extensive documentation. 
In addition to the fitting code itself, this release includes the original and filtered
experimental data, the fast NLO interpolation grids relevant for the
computation of hadronic observables, and whenever available
the bin-by-bin next-to-next-to-leading order (NNLO)
QCD and next-to-leading (NLO) electroweak $K$-factors for all processes entering the fit.
Furthermore, the code comes accompanied by a battery of plotting, statistical, and diagnosis tools
providing the user with an extensive characterisation of the PDF fit output.

The availability of the NNPDF open-source code, along with its detailed online
documentation, will enable users to perform new PDF
analyses based on the NNPDF methodology and modifications thereof.
Some examples of potential applications include
assessing the impact of new measurements in the global fit; producing
variants based on reduced datasets, carrying out PDF determinations with different theory
settings, e.g. as required for studies of $\alpha_s$ or heavy quark mass
sensitivity, or 
with different electroweak parameters; 
estimating the impact on the PDFs of theoretical constraints and calculations e.g. from
 non-perturbative QCD models~\cite{Ball:2016spl} or
 lattice calculations~\cite{Lin:2017snn,Cichy:2019ebf};
 and quantifying the role of 
theoretical uncertainties from missing higher orders to nuclear effects.
One could also deploy the NNPDF code as a toolbox to pin down the possible effects of beyond
the Standard Model physics at the LHC, such as Effective Field Theory corrections
in high-$p_T$ tails~\cite{Carrazza:2019sec, Greljo:2021kvv}
or modified DGLAP evolution from new BSM light degrees of freedom~\cite{Berger:2010rj}.
Furthermore, while the current version of the NNPDF code focuses on unpolarised parton distributions, its modular
and flexible infrastructure makes it amenable to the determination of closely
related non-perturbative collinear QCD quantities such as polarised PDFs, nuclear PDFs, fragmentation functions, or
even the parton distributions of mesons like pions and kaons~\cite{Adams:2018pwt}.

It should be noted that some of the functionalities
described above are already available within the
open source QCD fit framework {\tt xFitter}~\cite{Alekhin:2014irh,Zenaiev:2016jnq}.
The NNPDF code offers complementary functionalities
as compared to those in {\tt xFitter}, in particular by means of state-of-the-art
machine learning tools for the PDF parametrisation, robust methods for uncertainty
estimate and propagation, a wider experimental dataset,
an extensive suite of statistical validation and plotting tools, the possibility
to account for generic theoretical uncertainties, and an excellent
computational performance which makes possible full-fledged global PDF fits in less than one hour.

The main goal of this paper is to summarise the key features of the NNPDF code and
to point the interested reader to the online documentation, in which the code is presented in
detail and which, importantly, is kept up-to-date as the code continues to be developed
and improved.
First, in Sect.~\ref{sec:code}
we describe the structure of the code and its main
functionalities, including the relevant options.
The framework used to analyse the outcome of a PDF fit is described in
Sect.~\ref{sec:analysis}, while in Sect.~\ref{sec:applications}
we describe a few examples of possible applications for which users may wish to use the code.
We conclude and summarise some possible directions of future
development in Sect.~\ref{sec:conclusions}.

\section{Code structure}
\label{sec:code}
\label{sec:config}

The open-source {\tt NNPDF} framework enables performing global QCD
analyses of lepton-proton(nucleus) and proton-(anti)proton scattering data in
terms of the NNPDF4.0 methodology described in~\cite{nnpdf40}.
The code is publicly available from its {\sc\small GitHub} repository
\begin{center}
  {\tt \url{https://github.com/NNPDF/}}
\end{center}
and is accompanied by an extensive, continuously updated,
online documentation
\begin{center}
  {\tt \url{https://docs.nnpdf.science/}}
\end{center}
In this section, we describe the structure of the code and we present a
high-level description of its functionalities. We invite the reader to
consult the  documentation for details on its usage.\\[-0.3cm]

\begin{figure}
  \centering
\includegraphics[scale=.5]{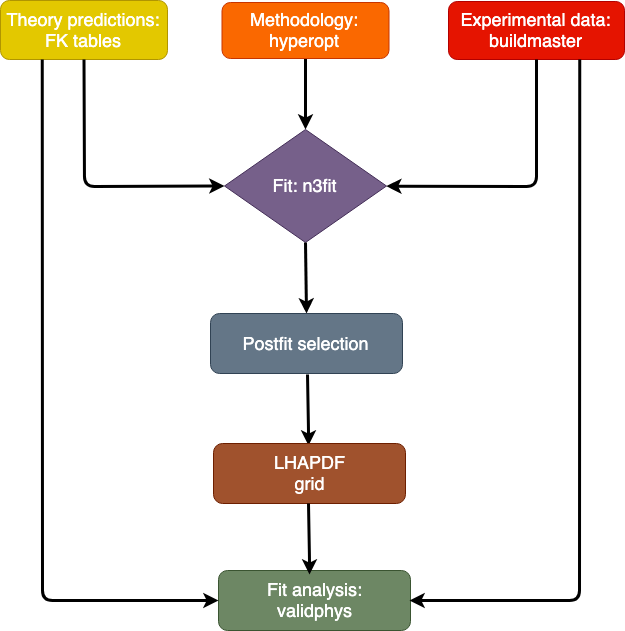}
\caption{\label{fig:wf} Schematic of the NNPDF code.
  The three main inputs are the theoretical calculations, encoded in terms
  of the precomputed {\tt FK}-tables, the methodological settings as determined by the hyperopt procedure,
  and the experimental data in the common {\tt buildmaster} format.
  The PDFs are fitted using {\tt n3fit}, and following a postfit
  selection the
  outcome is stored
  in the {\sc\small LHAPDF} grid format.
  Finally, a thorough characterisation of the results is carried out
  by the {\tt validphys} framework.
}
\end{figure}

The workflow for the {\sc\small NNPDF} code is illustrated in Fig.~\ref{fig:wf}.
The {\sc\small NNPDF} code is composed of the following main packages:
\begin{description}
 
\item[\textbf{The {\tt buildmaster} experimental data formatter}]
  A {\tt C++} code which transforms the original measurements provided
  by the experimental collaborations,
  e.g. via {\sc\small HepData}~\cite{Maguire:2017ypu},
  into a standard format that is tailored for PDF fitting.
  In
  particular, the code  allows for a flexible handling of experimental systematic
  uncertainties allowing
for different treatments of the correlated systematic uncertainties~\cite{Ball:2009qv,Ball:2012wy}.

    \vspace*{0.3cm}

  \item[\textbf{The {\tt APFELcomb} interpolation table generator}]
    This code takes hard-scattering partonic matrix element interpolators
    from {\tt APPLgrid}~\cite{Carli:2010rw} and {\tt
    FastNLO}~\cite{Wobisch:2011ij} (for hadronic processes) and {\tt
    APFEL}~\cite{Bertone:2013vaa} (for DIS structure functions) and combines
    them with the QCD evolution kernels provided by {\tt APFEL} to
    construct the fast interpolation grids called {\tt
    FK}-tables~\cite{Bertone:2016lga}. In this way, physical observables can
    be evaluated in a highly efficient manner as a tensor sum of {\tt
    FK}-tables with a grid of PDFs at an initial parametrisation scale $Q_0$.
    {\tt APFELcomb} also handles NNLO QCD and/or NLO electroweak
    $K$-factors when needed.\\
    
    Theory predictions can be generated configuring a variety of options,
    such as the perturbative order (currently up to NNLO), the values of the
    heavy quark masses, the electroweak parameters, the maximum number of
    active flavours, and the variable-flavour-number scheme used to account
    for the effects of the heavy quark masses in the DIS structure functions.
    The {\tt FK}-tables resulting from each choice are associated to a
    database entry trough a theory id, which allows to quickly identify them
    them.

    \vspace*{0.3cm}

  \item[\textbf{The {\tt n3fit} fitting code}]
    This code implements the fitting methodology described
    in~\cite{refId0,nnpdf40} as implemented in the {\tt TensorFlow}
    framework~\cite{abadi2016tensorflow}. The \texttt{n3fit} library allows
    for a flexible specification of the neural network model adopted to
    parametrise the PDFs, whose settings can be selected automatically via
    the built-in hyperoptimisation tooling ~\cite{2015CS&D....8a4008B}. These
    include the neural network type and architecture, the activation
    functions, and the initialisation strategy; the choice of optimiser and
    of its corresponding parameters; and hyperparameters related to the
    implementation in the fit of theoretical constraints such as PDF
    positivity~\cite{Candido:2020yat} and integrability. The settings for a
    PDF fit are input via a declarative run card. Using these
    settings, {\tt n3fit} finds the values of the neural network parameters,
    corresponding to the PDF at initial scale which describe the input data.
    Following a post-fit selection and PDF evolution step, the final output
    consists of an LHAPDF grid corresponding to the best fit PDF as well as
    metadata on the fit performance.
  
  \vspace*{0.3cm}

\item[\textbf{The {\tt libnnpdf} {\tt C++} legacy code}]
  A {\tt C++} library which contains common data structures together with
  the fitting code used to produce the NNPDF3.0 and NNPDF3.1 analyses~\cite{Ball:2014uwa,Ball:2016neh,Ball:2017nwa,Bertone:2017bmex,Ball:2017otu}.
  The availability of the {\tt libnnpdf} guarantees strict backwards
  compatibility of the NNPDF framework and the ability to benchmark the
  current methodology against the previous one.
  To facilitate the interaction between the NNPDF {\tt C++} and {\tt Python}
  codebases, we have developed {\tt Python} wrappers using the {\tt SWIG}~\cite{10.5555/1267498.1267513}
  library.

  \vspace*{0.3cm}

\item[\textbf{The {\tt validphys} analysis framework}] A package allowing to
analyse and plot data related to the NNPDF fit structures and I/O
capabilities to other elements of the code base.
The {\tt validphys} framework is  discussed in detail in
Sect.~\ref{sec:analysis}.

\end{description}

Complementing these main components, the {\sc\small NNPDF} framework also
contains a number of additional, ever-evolving,
tools which are described in the online documentation. 

\paragraph{Development workflow.}
The {\sc\small NNPDF} code adopts a development workflow compliant
with best practices in
professionally developed software projects.
Specifically, every code modification undergoes code review and is subjected to a suite of
automated continuous integration testing.
Moreover,
before merging into the main release branch, all relevant documentation is added
alongside any new tests that may be relevant to the incoming feature.
This feature
ensures that a broad code coverage within the test suite is maintained.

\paragraph{Installation.}
The various software packages that compose the NNPDF fitting code
can be installed via the binary packages provided by the {\tt conda}
interface, as described in
\begin{center}
  {\tt
    \url{https://docs.nnpdf.science/get-started/installation.html}
  }
\end{center}
The binary distribution allows users to
easily install the entire code suite alongside all relevant dependencies within
an isolated environment, which is also compatible with the one that has been tested
automatically.
Consequently, PDF fits can be produced with a known fixed version
of the code and all its dependencies, regardless of the machine where it is
running, hence ensuring the reproducibility of the result.
For the purposes of code development, it is also possible to set up an
environment where the dependencies are the same but the code can be edited,
allowing users to contribute to the open-source framework.


\paragraph{Input configuration.}

The settings that define the outcome
of a NNPDF fit are specified by means of a 
run card  written in {\tt YAML}, a common
human-readable data-serialisation language.
The main elements of fit run cards are:

\begin{description}
\item[\textbf{Input dataset:}] for each dataset, the user has to
  specify the NNPDF-internal string associated to it, the fraction of the data that
  goes into the training and validation subsets, and the inclusion of
  $K$-factors in the
  corresponding theoretical predictions.
  The latter are assigned different
  naming conventions depending on their nature: NNLO QCD, NLO electroweak,
  heavy-quark mass corrections for neutrino DIS~\cite{Gao:2017kkx}, or
  overall normalisation rescaling.
  Correlations between common systematic uncertainties between different datasets
  are automatically taken into account.
  
  \vspace*{0.2cm}

\item[\textbf{Kinematical cuts:}] a declarative format that specifies
    the cuts applied to the experimental data, based on the kinematics  of each data
    point and depending on the corresponding theory settings. 
    The cuts can be based on simple relations between the kinematics of each
    data point, such as the usual $Q^2_{\rm min}$ and $W^2_{\rm min}$ cuts
    applied to DIS structure functions, some derived quantity such as the
    value of the lepton charge asymmetry in $W$ decay data, or on more
    complex conditions such as retaining only points where the relative
    difference between NLO and NNLO predictions is below some threshold.
    These kinematical cut configuration can either be specified directly in the
    run card or the  built-in defaults can be used, and can be required for individual
    datasets or for types of processes instead.

      \vspace*{0.2cm}

\item[\textbf{Theory settings:}] the settings for theory predictions to be used in the fit,
   such as the
    perturbative order and the values of the coupling constants and of
    the quark masses, are specified an entry in the theory database,
    which in turn selects the set of
    {\tt FK}-tables, to be used during the fit.
    A wide range of {\tt FK}-tables for the most commonly used theory
    settings are already available and can be installed using the NNPDF code,
    while tables corresponding to different settings can also be assembled by
    the user whenever required. 
    The settings for the available entries of the theory database are
    specified in the online documentation.
    \vspace*{0.2cm}
    
  \item[\textbf{Fitting strategy and hyperparameters:}]
    the user can specify via the run card a number of methodological settings
    that affect the optimisation, such as the minimisation algorithm with the corresponding
    parameters, 
    the maximum training length, the neural network architecture and activation
    functions, and the choice of PDF fitting basis (e.g. using the evolution or
    the flavour basis).
    These methodological settings can either be set by hand or taken
    from the result of a previous {\tt hyperopt} run.
    Furthermore, random seeds can be configured to achieve different levels of correlation
    between Monte Carlo replicas across fits, as required e.g.
    for the correlated replica method used in the $\alpha_s(m_Z)$ extraction of ~\cite{Ball:2018iqk}.
    The user can additionally decide whether to save the weights of
    the neural networks during the fit or not, and whether to fit the Monte Carlo
    replicas or instead the central values of the experimental data.
    Another choice accessible via the run card is
    whether to use real data or instead fit to pseudo-data generated from a known underlying PDFs,
    as required during a closure test~\cite{Ball:2014uwa, closure40}.

    \vspace*{0.2cm}

  \item[\textbf{PDF positivity and integrability:}]
    as described in~\cite{nnpdf40}, in the NNPDF4.0 determination one imposes
    theoretical requirements on the positivity and integrability
    of the fitted PDFs by means of the Lagrange multiplier method.
    The user can then decide via the run card whether or not (or only partially) to impose
    these constraints on the PDFs, and if so define the initial values
    of the Lagrange multiplier weights.
    Note that some of the parameters governing the implementation of these theory
    requirements can also be adjusted by means of the hyperoptimisation procedure.
    
    \vspace*{0.2cm}
    
  \item[\textbf{Weighted fits:}]
    the user can  choose to give additional weight to specific datasets
    when computing the total $\chi^2$.
    This feature can be useful to investigate in more detail the relative impact
    that such datasets have in the global fit, and explore possible tensions
    with other datasets or groups of processes following the strategy laid out in~\cite{nnpdf40}.

\end{description}

The run cards required for producing the main NNPDF4.0 fits are stored under
\begin{center}
{\tt \url{https://github.com/NNPDF/nnpdf/tree/master/n3fit/runcards/reproduce_nnpdf40/}}
\end{center}
These enable users to readily reproduce the results and also generate modifications
of dataset selection, methodology or theory choices by suitably tweaking a run card.

\paragraph{Performance.}
One of the main advantages introduced by the new methodology
underlying NNPDF4.0 in comparison to its predecessors using genetic algorithms
is  the significant fitting  speed up achieved.
As an illustration of this improvement in performance, we note that
the NNPDF4.0 NNLO global fit takes fewer than
6 hours per replica on a single CPU core, as compared to \(\simeq 36\) hours using
the NNPDF3.1-like methodology.
This significant reduction of the CPU footprint
of the global PDF fits leads to a faster production rate of fit variants,
and it also allows one the prototyping of new approaches to PDF fitting using deep
 learning.
 Indeed, technologies such as hyperoptimisation were previously impractical
 but with the improved computational performance of the NNPDF code they are
 used in the fit.
 Furthermore, with the use of \texttt{TensorFlow} in the fitting toolkit, the
 ability to conveniently perform fits on the Graphics Processing Unit (GPU)
 might allow for further improvements in performance as suggested by the
 study in Ref.~\cite{Carrazza:2020qwu}.
 Such an
implementation in the main {\tt NNPDF} code is reserved for a future
release.

\section{The NNPDF analysis code: {\tt validphys}}
\label{sec:analysis}

The {\tt validphys} toolkit is at the heart of the NNPDF code base, bridging
together the other components and providing basic data structures,
compatibility interfaces, I/O operations and algorithms. These are used to 
assemble a suite of statistical analysis and plotting tools. We describe it
here, and refer the reader to the publications mentioned in the code structure
description in Sec.~\ref{sec:code} as well as the online documentation of the
NNPDF framework for further details on the other parts of the code.
 
\begin{figure}
  \centering
\includegraphics[width=0.99\textwidth]{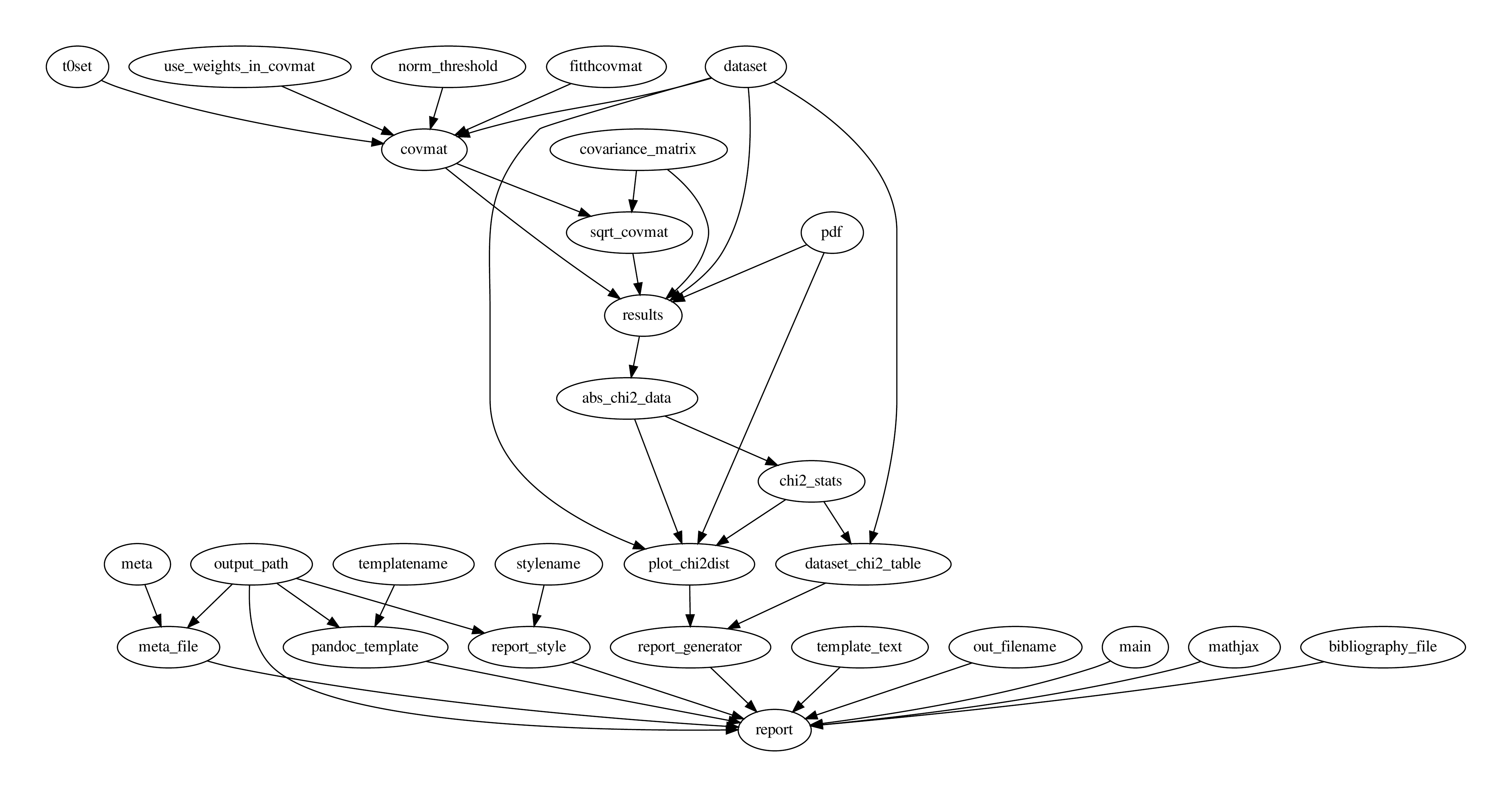}
\caption{\label{fig:graph} Direct acyclic graph corresponding to the
  run card provided in Fig.~\ref{fig:runcard}. The graph shows the
  inputs extracted from the run card, such as {\tt pdf} (the PDF set)
  and the {\tt dataset}, the intermediate steps
required for the $\chi^2$ computation (such as evaluating the {\tt covariance\_matrix}), and
the final target requested by the user, in this case a training {\tt report} containing a histogram
and a table with the $\chi^2$ values obtained for this dataset
for the indicated input PDF and choice of theory settings.}
\end{figure}

The {\tt validphys} code is in turn built on top {\tt
reportengine}~\cite{zahari_kassabov_2019_2571601}, a data analysis framework
which seeks to achieve the following goals:
\begin{itemize}
    \item To aid structuring data science code bases so as to make them
    understandable and lower the entry barrier for new users and developers.
    \item To provide a declarative interface that allows the user specifying the
    required analysis by providing a minimal amount of information in the
    form of a run card, making the analysis reproducible given said
    run card.
    \item To provide a robust environment for the execution of data analysis
    pipelines including robust error checking, automatic documentation,
    command-line tools and interactive applications.
\end{itemize}
The key observation underpinning the design of {\tt reportengine} is
that most programming tasks in data science correspond to codes that are
fully deterministic given their input.
Every such program can be seen as a
direct acyclic graph (DAG), see example the one shown in Fig.~\ref{fig:graph}, with links
representing the dependencies between a given step of the computation and the
subsequent ones.
Specifically, the nodes in such graph (resources)
correspond to results of executing functions (providers) which usually
correspond to functions in the {\sc\small Python} programming language.
These functions
are required to be pure, that is, such that their outputs are deterministic functions
of the inputs and that no side effects that alter the state of the program
happen.\footnote{Note that the concept of pure function is used here somewhat
more loosely than in programming languages such as {\tt
Haskell~\cite{mena2019practical}}, since side effects such as logging
information or writing files to disk are allowed as long as they are
idempotent.}
These side effects are
typically managed by the {\tt reportengine} framework itself, with tools to,
for example, save image files to a suitably unique filesystem location.

The goal of simplifying the programming structure is achieved then by
decomposing the program in terms of pure functions.
Code developers are required to reason
about the inputs of each individual function as well as its code, but not
about any global state of the program or the order of execution, with the
problem of putting the program together being delegated to the framework.

The {\tt reportengine} framework has extensive facilities for automatically
building the computation graph from the provided input.
Users
are only required to specify the ultimate target of the analysis (such as a
figure, table, or report) with the intermediate steps being deduced thanks to
a set of conventions in the program structure and a collection of utilities
provided by the framework (for example tools to implement the map-reduce pattern).
This allows complex analyses to be specified by
purely declarative run cards without the need to write custom code for each of
them.
In turn, the run cards allow
any user to precisely reproduce the results based on it and the corresponding
version of the code.

A simple {\tt validphys} run card, illustrating a minimal analysis of a
dataset is shown in Fig.~\ref{fig:runcard} with the DAG it spawns in
Fig.~\ref{fig:graph}.
%

As an example of the  meta-programming features of {\tt reportengine}, the
{\tt template\_text} input in the runcard displayed in Fig.~\ref{fig:runcard} illustrates how it is
possible to spawn arbitrary other actions, with their corresponding
dependencies, based on the user input as shown in Fig.~\ref{fig:graph}.
The framework allows implementing similar complex workflows with its
programming interface.
Users are referred to the online documentation for
further details, code references, and specific examples.
%

\begin{figure}
  \centering
  \begin{minipage}{0.5\textwidth}
  \begin{framed}
  \begin{large}
\begin{Shaded}
\begin{Highlighting}[]
\FunctionTok{dataset\_input}\KeywordTok{:}
\AttributeTok{    }\FunctionTok{dataset}\KeywordTok{:}\AttributeTok{  ATLAS\_WP\_JET\_8TEV\_PT}
\AttributeTok{    }\FunctionTok{cfac}\KeywordTok{:}\AttributeTok{ }\KeywordTok{[}\AttributeTok{QCD}\KeywordTok{]}

\FunctionTok{theoryid}\KeywordTok{:}\AttributeTok{ }\DecValTok{200}

\FunctionTok{use\_cuts}\KeywordTok{:}\AttributeTok{ }\StringTok{"nocuts"}

\FunctionTok{pdf}\KeywordTok{:}\AttributeTok{ NNPDF31\_nnlo\_as\_0118}

\FunctionTok{template\_text}\KeywordTok{: }\CharTok{|}
\NormalTok{    \# Histogram}

\NormalTok{    \{@plot\_chi2dist@\}}

\NormalTok{    \# Table}

\NormalTok{    \{@dataset\_chi2\_table@\}}


\FunctionTok{actions\_}\KeywordTok{:}
\AttributeTok{    }\KeywordTok{{-}}\AttributeTok{ report(main=True)}
\end{Highlighting}
\end{Shaded}
\end{large}
\end{framed}
\end{minipage}
\caption{\label{fig:runcard} A {\tt validphys} runcard which produces a
  report containing a table and a histogram with the $\chi^2$ values 
  obtained for the ATLAS $W^++{\rm jets}$ 8 TeV differential
  distributions when using
  the $N_{\rm rep}=100$ replicas of NNPDF3.1 NNLO as input dataset and the theory settings specified by
  the {\tt theoryid: 200} of the database.
  In particular, the runcard specifies the string for the {\tt dataset},
  the use of QCD $K$-factors, and the requirements that no kinematic cuts
  should be applied to the input dataset.
  Possible input options are referenced in Sec.~\ref{sec:config}.
  The DAG graph corresponding to the execution of this runcard is
  represented in Fig.~\ref{fig:graph}.
 }
\end{figure}

The introspection capabilities of {\tt reportengine} enable it to provide a
robust and convenient environment for carrying out analysis. Most notably
they enable specifying powerful checks on the user input.
Basic constraints are implemented by
instrumenting type annotations of {\sc\small Python} functions, which are used to verify
that data types in the run cards match those expected by the code, but in
addition arbitrary checks can also be attached to both input values or
provider functions. This is commonly known as contract programming, but
differs with many implementations in that checks are executed at the time the
DAG is being built instead of when functions are executed. Thus, the DAG
construction phase can be seen as a compilation phase, where developers have
the ability to write arbitrary compiler checks. This feature allows
eliminating large classes of runtime errors, thereby increasing the chances
that the analysis runs to completion once the DAG has been constructed and
checked. 
Another introspection feature consists of the capability of tracing the
required inputs for a given provider and displaying them as automatically
generated command line documentation.

As an implementation of the {\tt reportengine} framework the {\tt validphys}
code features workflow focused on declarative and reproducible run cards. The
code relies on common {\tt Python} data science libraries
such as {\tt NumPy}~\cite{harris2020array}, {\tt
SciPy}~\cite{2020SciPy-NMeth}, {\tt Matplotlib}~\cite{4160265} and {\tt
Pandas}~\cite{mckinney-proc-scipy-2010} through its use of {\tt
Pandoc}~\cite{macfarlane2013pandoc}, and it implements data structures that
can interact with those of {\tt libnnpdf}, as well as with analogs written in
pure {\sc\small Python}.
These include NNPDF
fits, {\tt LHAPDF} grids, and {\tt FK}-tables.
In addition, the code allows to quickly acquire relevant data and theory
inputs by automatically downloading them from remote sources whenever they
are required in a runcard.
It also contains tooling to upload analysis
results to an online server, to share it with other users or developers, 
and to allow it to be reproduced by other parties.

Some common data analysis actions that can be realised within the {\tt validphys}
framework include:
\begin{itemize}
\item Evaluating the convolutions between {\tt FK}-tables and PDF sets, to evaluate
  in a highly efficient manner the theoretical predictions for the cross-sections
  of those datasets and  theory settings we have implemented.
  Note that here any input PDF set can be used, not only NNPDF sets.
  
\item Producing data versus theory comparison plots allowing for
  the graphical visualisation of the wealth of
    experimental measurements implemented in the NNPDF framework  matched against the
    theory predictions.
    Again, predictions for arbitrary PDF sets can be used as input.
    
    \item Computing  statistical estimators based on such data versus theory comparison,
    such as the various types of $\chi^2$~\cite{Ball:2009qv}, together
    with many plotting and grouping options.
    \item A large variety of plotting tools and options for the PDFs and partonic luminosities,
      including plots in arbitrary PDF bases.
      Some of these functionalities are related to those provided by the {\tt APFEL-Web}
      online PDF plotter~\cite{Carrazza:2014gfa}.
\item Manipulating LHAPDF grids, implementing operations such as Hessian
conversions~\cite{Carrazza:2015aoa,Carrazza:2016htc}.
\end{itemize}

\begin{figure}
  \centering
    \includegraphics[width=.85\textwidth]{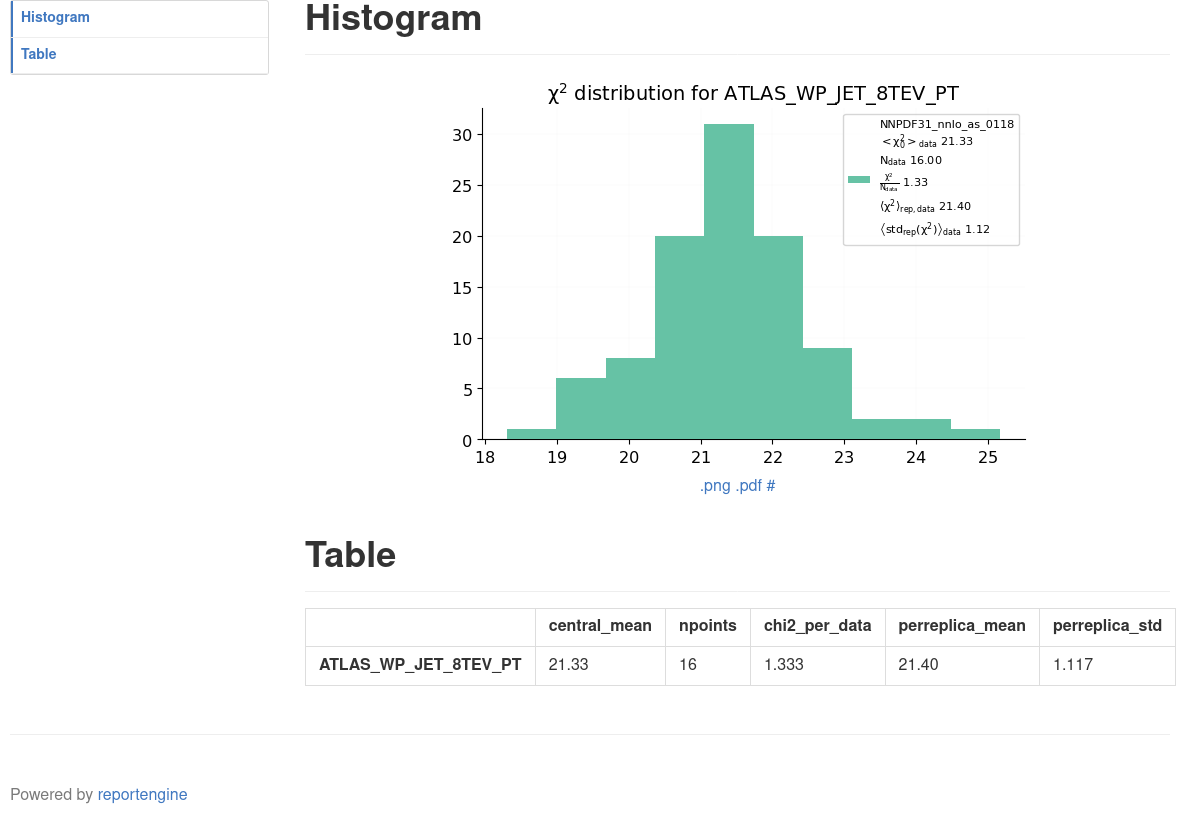}
    \caption{\label{fig:report} The output of executing the runcard in
      Fig.~\ref{fig:runcard} with {\tt validphys} is an
      HTML report consistent of an histogram and
      the corresponding table indicating the distribution of $\chi^2$ values
      over the $N_{\rm rep}=100$ replicas of NNPDF3.1 NNLO for the
      ATLAS $W^++{\rm jets}$ 8 TeV differential
distributions~\cite{Aaboud:2017soa} and the {\tt theoryid:200} theory settings.}
\end{figure}

The typical output of {\tt validphys} is an HTML
report containing the results requested by the user
via the runcard.
Fig.~\ref{fig:report} displays the report obtained after executing the runcard in
Fig.~\ref{fig:runcard}, consistent of an histogram displaying the
distribution of $\chi^2$ values for the $N_{\rm rep}=100$ replicas
of the NNPDF3.1 NNLO set when its predictions based on the {\tt theoryid:200} theory settings
are compared to the  ATLAS $W,Z$ 13 TeV
total cross-sections.
In order to highlight the potential of {\tt validphys}, we have collected in this link
\begin{center}

{\tt \url{https://data.nnpdf.science/nnpdf40-reports/}}
  
\end{center}  
representative training reports corresponding to the NNPDF4.0 analysis, such
as comparisons between fits at different perturbative orders and between
fits based on different datasets.

Additional features of the current
release of the {\tt validphys} framework include tools that make possible:
\begin{itemize}
    \item Comparing two PDF fits by means of the  {\tt vp-comparefits} tool, which generates a report
      composed by almost 2000 figures and and 12 tables, displaying
      fit quality estimators, PDF comparisons, data-theory
    comparisons and positivity observables.
  \item Carrying out and characterising closure tests~\cite{closure40}
    and future tests~\cite{Cruz-Martinez:2021rgy}.
    \item Performing  simultaneous fits of the PDFs together with the strong
      coupling constant~\cite{Ball:2018iqk}.
    \item Evaluating the theory covariance matrix constructed from scale variations,
      which can then be used as input for PDF fits accounting for missing higher order uncertainties (MHOUs)
      following the strategy of~\cite{AbdulKhalek:2019ihb,AbdulKhalek:2019bux}.
    \item Studying Hessian PDF tolerances.
    \item Determining Wilson coefficients in the Effective Field Theory (EFT) framework together with PDFs following the strategy presented in~\cite{Carrazza:2019sec, Greljo:2021kvv}.
    \item  Analysing theoretical prediction with matched scale variations.
\end{itemize}

In conclusion, it is worth emphasising that many of the {\tt validphys} features
described here
can be deployed outside the framework of the NNPDF fits.
For instance, the tooling to evaluate the theory covariance matrix could also be relevant in the context
of Hessian PDF fits, and comparisons between PDF sets can be carried out for other
fits beyond NNPDF, provided one is careful and adopts consistent theoretical
settings for each of the inputs.

\section{Applications}
\label{sec:applications}

Let us briefly discuss now some possible future applications
of the NNPDF fitting framework presented in this work.
As representative examples, we consider the 
inclusion of new experimental data,
producing fits varying the theory settings, and
going beyond the determination unpolarised collider PDFs.
We discuss each of these applications in turn, and
for a more comprehensive list
we refer the interested user to the online documentation.

\paragraph{Adding new experimental data to the global fit.}
A typical application of the open-source NNPDF framework would be that
of assessing the
impact of some new measurement in the global PDF fit.
Carrying out a full-fledged fit has several advantages as compared
to approximate methods such as Bayesian reweighting~\cite{Ball:2011gg,Ball:2010gb}, in particular one does
not rely on the availability of prior fits composed by a very large number of replicas.
Also, this way the user can easily vary the input dataset or the theoretical settings of this
baseline fit.
Furthermore, it is possible to add simultaneously to the global fit a large number
of new datasets, while the reliability of the reweighting procedure
is typically limited to a single dataset.

To implement a new dataset in the NNPDF code, one should start
by adding the new measurement to the {\tt buildmaster} suite.
This will parse the new data points into the common format
suitable for its use in the PDF fits.
Such an implementation will in general include
information regarding the data central values, specification of the kinematic
variables, statistical uncertainties and any relevant correlated systematic uncertainties
that may exist in the dataset.
In particular, the systematic uncertainties must
be accompanied by metadata specifying their type (i.e if they are multiplicative
or additive) as well as any possible correlations they may have with other
systematic uncertainties (for example, the luminosity uncertainty will often
be correlated across a given experiment).
These uncertainties will then be used
to construct the covariance matrix as well as the Monte Carlo replicas
used to train the neural networks parametrising the PDFs.

Furthermore, in order to run the fit,
the user would have to produce the corresponding {\tt FK}-tables
for this new dataset, which implies evaluating the fast NLO grids
via {\tt APPLgrid}, {\tt FastNLO}, or {\tt PineAPPL}~\cite{Carrazza:2020gss} and then combining
them with the DGLAP evolution kernels via {\tt APFELcomb}. 
Depending on the perturbative order and electroweak settings of the fit,
one needs to complement this {\tt FK}-table with bin-by-bin NNLO QCD and/or NLO
electroweak $K$-factors.
With these ingredients, it is then be possible to add the data to a NNPDF fit
and gauge its impact by comparing to a baseline with this same dataset excluded.
If the impact of the dataset on the PDFs is moderate one
can adopt the same hyperparameters as in the baseline reference; however, it
is recommended practice to verify the stability of the fit results
with respect a dedicated round of  hyperoptimisation.
Note also that new theory constraints, e.g. as those that could be imposed
by lattice QCD calculations, can be accounted for in the same manner
is with the addition of a new dataset.

As a consequence of adding the new dataset to those already packaged within the
{\sc\small NNPDF} code, the user now has access to the \texttt{validphys}
tools described in Sect.~\ref{sec:analysis}, and hence they can easily
characterise the goodness of the fit and quantify the agreement
with the theory predictions, and well as assess the impact of this new dataset
into the PDFs, partonic luminosities, and physical cross-sections.

\paragraph{NNPDF fits with different theory settings.}
Another foreseeable application of the open source fitting code
is to produce variants of the NNPDF global analyses
with modifying settings for the theoretical calculations.
For instance, in determinations of the strong coupling $\alpha_s(m_Z)$ from
collider data, one typically needs a series of PDF fits with a wide range and fine
spacing of $\alpha_s$ values.
These dedicated fits can be produced with the NNPDF code,
and in addition while producing such PDF fits the user can also choose to tune
the input dataset, e.g. by excluding specific types of processes, and
the theory settings, e.g. with different variable-flavour-number-scheme.
As emphasised in~\cite{Forte:2020pyp}, when extracting SM parameters such as $\alpha_s(m_Z)$
from datasets sensitive to PDFs, it it necessary to simultaneously
account for the impact of such datasets on the PDFs themselves (and not only on $\alpha_s$) to avoid
biasing the determination.
Hence, these varying-$\alpha_s$ fits should already include the dataset from which
$\alpha_s$ will be extracted, and this is only possible thanks to the availability
of the NNPDF open source code.

The same caveats apply in the case of determinations of the heavy quark (charm, bottom, and top)
masses from collider processes in which PDFs also enter the theory calculations.
Other possible examples of NNPDF fits with varying theory settings are fits
with different flavour assumptions, DGLAP evolution settings, or with approximations
for unknown higher order perturbative corrections such as those evaluated from
resummation.
One may also be interested in tailored PDF sets for specific cross-section
calculations, such as the doped PDFs~\cite{Bertone:2015gba} where the running with the active
number of flavours $n_f$
is different for $\alpha_s(Q)$ and for the PDF evolution.

In order to run a variant of the NNPDF fit with different theory settings,
the user needs to verify if the corresponding sought-for {\tt theory-id} already exists in
the {\tt theory} database.
If this is the case, the fit with the new theory settings can be easily produced
by adjusting the {\tt theory-id} parameter in the run card.
If, however, the {\tt FK}-tables with the required theory settings
are not available in the database, the user needs first to produce them
using {\tt APFELcomb}.
We note that this is a relatively inexpensive step from the computational point
of view, provided the corresponding NLO fast grids and the associated  $K$-factors have
been already produced.
The user can follow the instructions in
\begin{center}
{\tt \url{https://docs.nnpdf.science/tutorials/apfelcomb.html}}
\end{center}
to produce {\tt FK}-tables with their desired settings
and assign them to a new  {\tt theory-id} in the theory database.
By means of the {\tt validphys} tooling, this new set of {\tt FK}-tables can also
be uploaded to the theory server where it will become available for other users.

\paragraph{Beyond unpolarised collinear PDFs.}
The current version of the NNPDF code focuses on unpolarised parton distributions.
However, its flexible and modular infrastructure can be extended to
the determination of related non-perturbative QCD quantities by means
of the same methodology.
While the NNPDF approach has also been used for the determination
of polarised PDFs~\cite{Nocera:2014gqa,Ball:2013lla},
fragmentation functions~\cite{Bertone:2017tyb,Bertone:2018ecm}, and
nuclear PDFs~\cite{AbdulKhalek:2019mzd,AbdulKhalek:2020yuc},
in all these cases the code infrastructure only partially overlaps
with that underlying NNPDF4.0.
For instance, the polarised PDF determination rely on the {\sc\small Fortran} predecessor
of the NNPDF code, while the nuclear PDF fits adopt the {\tt FK}-table
approach for theoretical calculations but are based on a stand-alone machine learning
framework.
The availability of the NNPDF framework as open source code should hence lead to
progress into its extension to other quantities beyond unpolarised collinear PDFs,
as well as for the determination of the collinear PDFs of different
hadronic species such as pions or kaons.
These studies are especially interesting at the light of future experiments
with focus on testing the nucleon, nuclear, and mesonic structure,
from the Electron Ion Colliders~\cite{Anderle:2021wcy,AbdulKhalek:2021gbh}
to AMBER at the CERN-SPS~\cite{Adams:2018pwt}.

A closely related application of the NNPDF fitting code would be
the simultaneous determination of non-perturbative QCD quantities exhibiting
non-trivial cross-talk,
such as nucleon and nuclear PDFs~\cite{Khalek:2021ulf},  (un)polarised PDFs together with
fragmentation functions~\cite{Moffat:2021dji}, or collinear and transverse-momentum-dependent
PDFs.
Such integrated global PDF determinations have many attractive features, for instance
in the proton global analysis it would not be necessary anymore to treat in a special
manner the deuteron and heavy nuclear datasets (since the $A$ dependence
would be directly extracted from the data), and the interpretation of processes such as semi-inclusive
DIS (SIDIS) would not rely on assumptions about the behaviour of either the nucleon
PDFs (for the initial state) or the fragmentation functions (for the final state).
Clearly, a pre-requisite for such integrated fits is the availability of
the code infrastructure for the determination of the individual non-perturbative QCD quantities
within the public NNPDF framework.

\section{Conclusions}
\label{sec:conclusions}

In this work we have presented the public release, as an open-source
code, of the software framework underlying the recent NNPDF4.0
global determination of parton distributions.
The flexible and robust NNPDF  code exploits  state-of-the-art
developments in machine learning to realise a comprehensive
determination of the proton structure from a wealth of experimental data.
The availability of this framework as open source should encourage
the broader high-energy and nuclear physics communities to deploy machine
learning methods in the context of PDF studies.

Among the wide range of possible user cases provided by the NNPDF code, one can list
assessing the impact of new data, producing tailored fits
with variations of SM parameters such as $\alpha_s(m_Z)$ or $m_c$ for their
simultaneous extraction together with the PDFs,
and studying the eventual presence of beyond the SM physics in precision
LHC measurements of the high-$p_T$ tails of kinematic distributions
using effective field theories.
Furthermore, the determination of  related non-perturbative QCD quantities from nuclear PDFs
and polarised PDFs to fragmentation functions represents another potential
application of the NNPDF framework

In order facilitate these various applications, the NNPDF codebase is now almost entirely written in
{\sc\small Python}, the currently \textit{de facto} standard choice of programming
language within both the data science as well as the scientific community.
With the majority of the libraries being highly efficient wrappers of faster
languages, {\sc\small Python} is no longer bottle-necked by performance and so
its relatively low barrier of entry should allow for the NNPDF code to be
modified and expanded.
With this motivation, we have discussed how the user may wish to configure a run card for their PDF fit,
indicated the details of the parameters that are exposed to the user,
and presented the \texttt{validphys}
library which acts as an in-house analysis suite designed to be not only reproducible, but
also allowing for complex tasks to be achieved using transparent run card based inputs.

We reiterate that we have restricted ourselves
to a succinct high-level summary
of the main functionalities of the {\sc\small NNPDF} code.
The main reference for the interested user
is online documentation which accompanies
this release,
 which features technical commentary as well as
 example use cases.
 The documentation is kept
 continuously up-to-date following the ongoing development of the code.

\section*{Acknowledgements}

S.~C., S.~F., J.~C.-M., R.~S. and C.~S. are supported by the European
Research Council under the European Union’s Horizon 2020 research and
innovation Programme (grant agreement n.740006).
M.~U. and Z.~K. are supported by the European Research Council under the European Union’s
Horizon 2020 research and innovation Programme (grant agreement
n.950246). M.~U. and S.~I. are partially supported by the Royal Society grant
RGF/EA/180148. The work of M.~U. is also funded by the Royal Society
grant DH150088. The work of M.~U., S.~I., C.~V. and Z.~K. is partially supported
by the STFC consolidated grant ST/L000385/1. The work of Z.~K. was partly
supported by supported by the European Research Council Consolidator Grant
“NNLOforLHC2”.  J.~R. is partially supported by NWO, the Dutch
Research Council.
C.~V. is supported by the STFC grant ST/R504671/1.
T.~G. is supported by The Scottish Funding Council, grant H14027.
R.~L.~P.  and M.~W. by the STFC grant ST/R504737/1.
R.~D.~B., L.~D.~D. and E.~R.~N. are supported by the STFC grant
ST/P000630/1.
E.~R.~N. was also supported by the European
Commission through the Marie Sklodowska- Curie Action ParDHonS (grant
number 752748). 

\providecommand{\href}[2]{#2}\begingroup\raggedright\endgroup

\end{document}